\documentclass[aps,prl,reprint,showpacs,superscriptaddress]{revtex4-1}
\usepackage{amssymb}
\usepackage{amsmath}
\usepackage[dvips]{graphicx}

\usepackage{hyperref}
\hypersetup{
colorlinks=true, 
breaklinks=true, 
urlcolor= blue, 
linkcolor= red, 
bookmarksopen=false, 
}

\def\ket#1{\mathinner{|{#1}\rangle}}

\def\upp{\uparrow}
\def\dnn{\downarrow}

\renewcommand{\imath}{\mathrm{i}}

\newcommand{\lcRate}{\gamma_{c}}
\newcommand{\exRateTh}{\omega_{MF}}
\newcommand{\exRate}{\omega_{\textit{ex}}}

\usepackage[usenames]{color}

\begin{document}

\title{Spin waves and Collisional Frequency Shifts of a Trapped-Atom Clock }

\author{Wilfried Maineult}
\affiliation{LNE-SYRTE, Observatoire de Paris, CNRS, UPMC, 61 av
de l'Observatoire, 75014 Paris, France}

\author{Christian Deutsch}
\affiliation{Laboratoire Kastler Brossel, ENS, UPMC, CNRS, 24 rue Lhomond, 75005 Paris, France}

\author{Kurt Gibble} \affiliation{Department of Physics, The Pennsylvania State
University,  Pennsylvania 16802, USA}

\author{Jakob Reichel}
\affiliation{Laboratoire Kastler Brossel, ENS, UPMC, CNRS, 24 rue Lhomond, 75005 Paris, France}

\author{Peter Rosenbusch}%
\email{Peter.Rosenbusch@obspm.fr}
\affiliation{LNE-SYRTE, Observatoire de Paris, CNRS, UPMC, 61 av
de l'Observatoire, 75014 Paris, France}

\date{\today}

\begin{abstract}
We excite spin-waves with spatially inhomogeneous pulses and study the resulting frequency shifts of a chip-scale atomic clock of trapped $^{87}$Rb. The density-dependent frequency shifts of the hyperfine transition simulate the s-wave collisional frequency shifts of fermions, including those of optical lattice clocks. As the spin polarizations oscillate in the trap, the frequency shift reverses and it depends on the area of the second Ramsey pulse, exhibiting a predicted beyond mean-field frequency shift. Numerical and analytic models illustrate these observed behaviors.
\end{abstract}

\maketitle
Quantum scattering in an ultracold gas of indistinguishable spin-1/2
atoms leads to rich and unexpected behaviors, even above the onset of
quantum degeneracy. Among these, spin waves are a beautiful
macroscopic manifestation of identical spin rotation (ISR)
\cite{LaloeISRE1982,JohnsonPRL52_1508,LewandowskiSpinWaves2002,DuSpinWavesFermions2009}.  ISR also inhibits dephasing, which can dramatically increase
the coherence time in a trapped ensemble of interacting atoms to tens
of seconds \cite{TACCISRE2010,Kleine11}, with possible applications to
compact atomic clocks and quantum memories.  Another example is
collisional interactions in optical lattice clocks
\cite{Campbell2009,GibbleShift2009,ReyFermionShift2009,Yu2010}. Their detailed
understanding is a prerequisite for optical lattice clocks to realize their full potential as future primary standards.

At ultracold temperatures, scattering is purely s-wave, which is
forbidden for indistinguishable fermions, suggesting that clocks using
ultracold fermions are immune to collision shifts \cite{Gibble1995,Zwierlein2003}.  However, spatial inhomogeneities of the
clock field, which are naturally larger for optical frequency fields
than for radio-frequencies, allow fermions to become distinguishable
and therefore can lead to s-wave clock shifts \cite{GibbleShift2009,ReyFermionShift2009,Yu2010}. A series of experiments attributed the collision
shifts of Sr lattice clocks to these novel s-wave fermion collisions
with inhomogeneous clock field excitations \cite{Campbell2009,
SwallowsSupprCollShift2011,Bishof2011}. However, subsequent work
showed that p-waves dominate for Yb lattice clocks, and p-wave
scattering is consistent with all the observed Sr collisional
frequency shifts \cite{Ludlow2011}.

Bosons with state-independent scattering lengths have fermion-like
exchange interactions
\cite{FuchsSpinWaves2002,GibbleShift2009,GibbleViewpoint2010}.  This allows us
to simulate  the s-wave fermion collisional shift with a
chip-scale clock that traps $^{87}$Rb, a boson with nearly equal scattering lengths.
 We observe the distinguishing feature that the collisional
shift in the presence of inhomogenous excitations depends on the area
of the second Ramsey clock pulse. This dependence sets it apart from
the well-known s-wave shift for homogeneous excitations,
which is absent for fermions and, for bosons, depends
only on the first pulse area, and hence the population dif-
ference of the two clocks states \cite{Harber2002,GibbleShift2009}. Further, inhomogeneous excitations
directly excite spin waves. We show an inextricable link between
spin-waves and the s-wave fermion collisional shifts. Notably, we observe
frequency shifts that change sign as spin polarizations oscillate
in the trap.

We perform Ramsey spectroscopy with two spatially inhomogeneous pulses
to study spin waves and the collisional frequency shifts of trapped $
^{87}$Rb atoms.  The first clock pulse creates an inhomogeneous spin polarization, which varies linearly in space. We directly observe a spatiotemporal oscillation
of this spin polarization, which characterizes the strength of the
atomic interactions. Driving a second Ramsey pulse, we measure
frequency shifts of this clock. Here we vary the areas of each
pulse and the interrogation time between the two pulses, to probe the
unique behaviors of s-wave fermion clock shifts. We develop
analytic and numerical models that describe the observed spin-waves and the novel dependence on the area of the second Ramsey pulse.

\begin{center}
\begin{figure}
\includegraphics[width=1.00\linewidth,trim=0.5cm 0.9cm 0 1.4cm,clip]{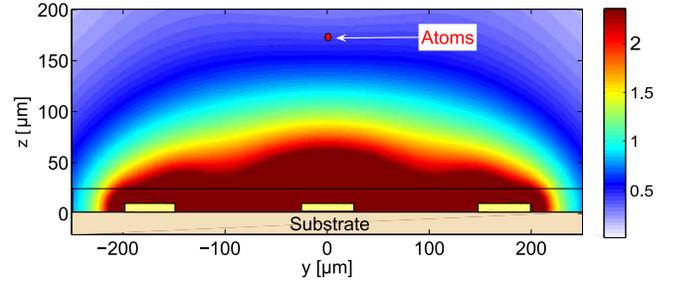}
\caption{Calculated microwave field amplitude of the coplanar waveguide on our atom chip, in arbitrary units. The rapidly decaying near-field causes a small vertical gradient of the spin-polarization across the trapped atom cloud.} \label{Fig:MWfield}
\end{figure}
\end{center}

Our chip-scale atomic clock magnetically traps between $10^3$ and $10^5$ atoms at a distance $z_0 = 156\,\mu$m below a microwave coplanar waveguide 
on our atom chip \cite{TACC2010,TACCISRE2010}. A microwave and radio-frequency, two-photon excitation drives the clock transition, $\ket{\dnn}\equiv\ket{F=1,m_F=-1}$ to $\ket{\upp}\equiv\ket{F=2,m_F=1}$. The near field of the microwave guide (Fig. \ref{Fig:MWfield}) creates a slightly inhomogeneous Rabi frequency in the vertical $z$ direction, $\Omega(\mathbf{r}) = \Omega_0(1 + \delta_1 z + \delta_2 z^2+ \ldots)$.
The trap frequencies are $(\omega_x,\omega_y,\omega_z) =2\pi (32(1),97(1),119.5(5))$ Hz.
The temperature of the cloud is 175(6) nK, at least 30 nK above the onset of Bose Einstein condensation, with no measurable dependence on the atom number.
At this temperature, the identical spin rotation rate $\exRate$, of order  $ \exRateTh\equiv4\pi \hbar |a_{\upp\dnn}| \bar{n}/m $, dominates in our experiments. The lateral collision rate $\lcRate\propto a_{\upp\dnn}^2\bar{n}v_T$  is always much lower than the trap frequencies, corresponding to the Knudsen regime. Here $a_{\upp\dnn}$ is the inter-state scattering length, $\bar{n}$ the mean density, $m$ the atomic mass and $v_T$ the thermal velocity. The magnetic field at the trap center is tuned to minimize the inhomogeneous spread of transition frequency \cite{TACCPeter2009} so that dephasing can be neglected on the timescales we consider \footnote{The inhomogeneous spread of transition frequencies due to the magnetic potential and the spatially varying atom density, is less than 80~mHz.}.

\begin{center}
\begin{figure}
\includegraphics[width=.25\linewidth,trim=0.65cm .15cm .65cm -1cm,clip]{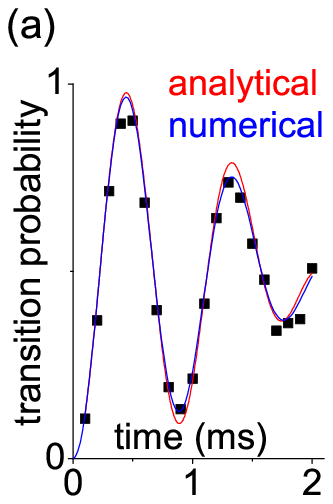}
\includegraphics[width=.7\linewidth,trim=0.3cm 1cm 0cm 0cm,clip]{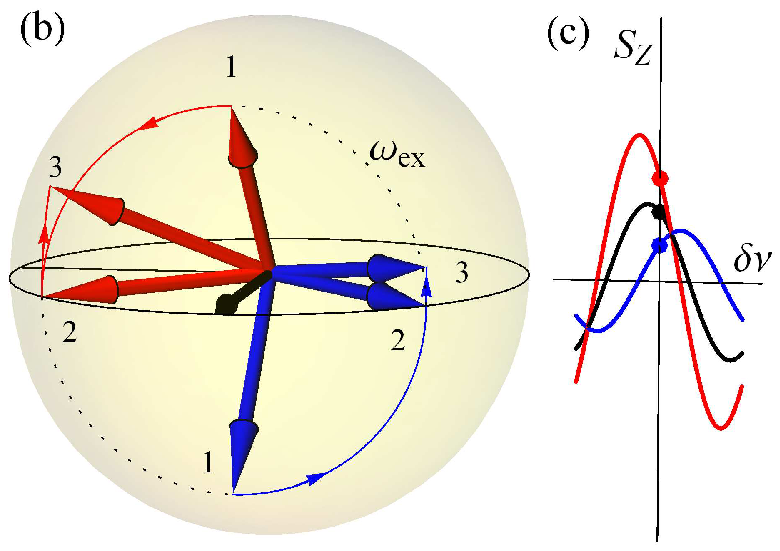}
\caption{(a) Inhomogeneous Rabi flopping with $\Omega(\mathbf{r}) = \Omega_0(1 + \delta_1 z)$. The data are fitted to models of two trapped atoms: (red) an analytic model with a single motional sideband gives $\delta_1 = 0.147(4)~\xi_z^{-1}$. (blue) numerical simulation for representative vibrational states $\eta_z$ = $30\pm8$ yields $\delta_1 = 0.090(4)~\xi_z^{-1}$. (b) Bloch sphere evolution of the spins of two representative atoms (blue and red), which are initially on opposite sides of the trap in state $\ket{\dnn}$. The first inhomogeneous Ramsey excitation pulse rotates the spins differently ($\theta_{red,1}=4\pi/5$ and $\theta_{blue,1}=\pi/5$). During the Ramsey interrogation time, the scattering produces an ISR rotation of the two spins around their sum (black arrow), here, by $\omega_{ex}T_R=\pi/2$. We take a second pulse with the same inhomogeneity as the first, but weaker, $\overline{\theta_2}=\pi/8$. It barely moves the blue spin, whereas the red rotates to be more vertical. We detect the vertical projection of each spin $S_z$, corresponding to the points in (c). In (c) we trace $S_z$ as a function of detuning, showing the resonance shifted to a negative detuning. For $\omega_zT_R=(2j+1)\pi$, both atoms have switched sides of the trap so that the inhomogeneity of the 2nd pulse instead gives the blue spin a larger rotation, hence higher Ramsey fringe contrast. Thus the frequency shift changes sign as the spins oscillate in the trap. } \label{Fig:BlochVectors}
\end{figure}
\end{center}

The variation of the  Rabi frequency across the atom cloud
is determined by fitting the resonant Rabi flopping using $\Omega_0\gg\omega_z$ so that atomic motion during the pulse can be neglected (Fig. \ref{Fig:BlochVectors}(a)). We find $\delta_1\approx0.1~\mathrm{\xi_z}^{-1}$ and $\delta_2\approx0$ which is reasonable since the r.m.s. cloud radius $\xi_z= 4.1\mathrm{\mu m}\ll z_0 $. \footnote{We checked experimentally that the inhomogeneity is predominantly vertical. Along $y$ there is no detectable variation of $\Omega(\mathbf{r})$ and its variation along $x$ is 5 times smaller than along $z$}.

\begin{center} \begin{figure}
\includegraphics[width=1.0\linewidth,trim=0 0.5cm 1cm 12cm,clip]{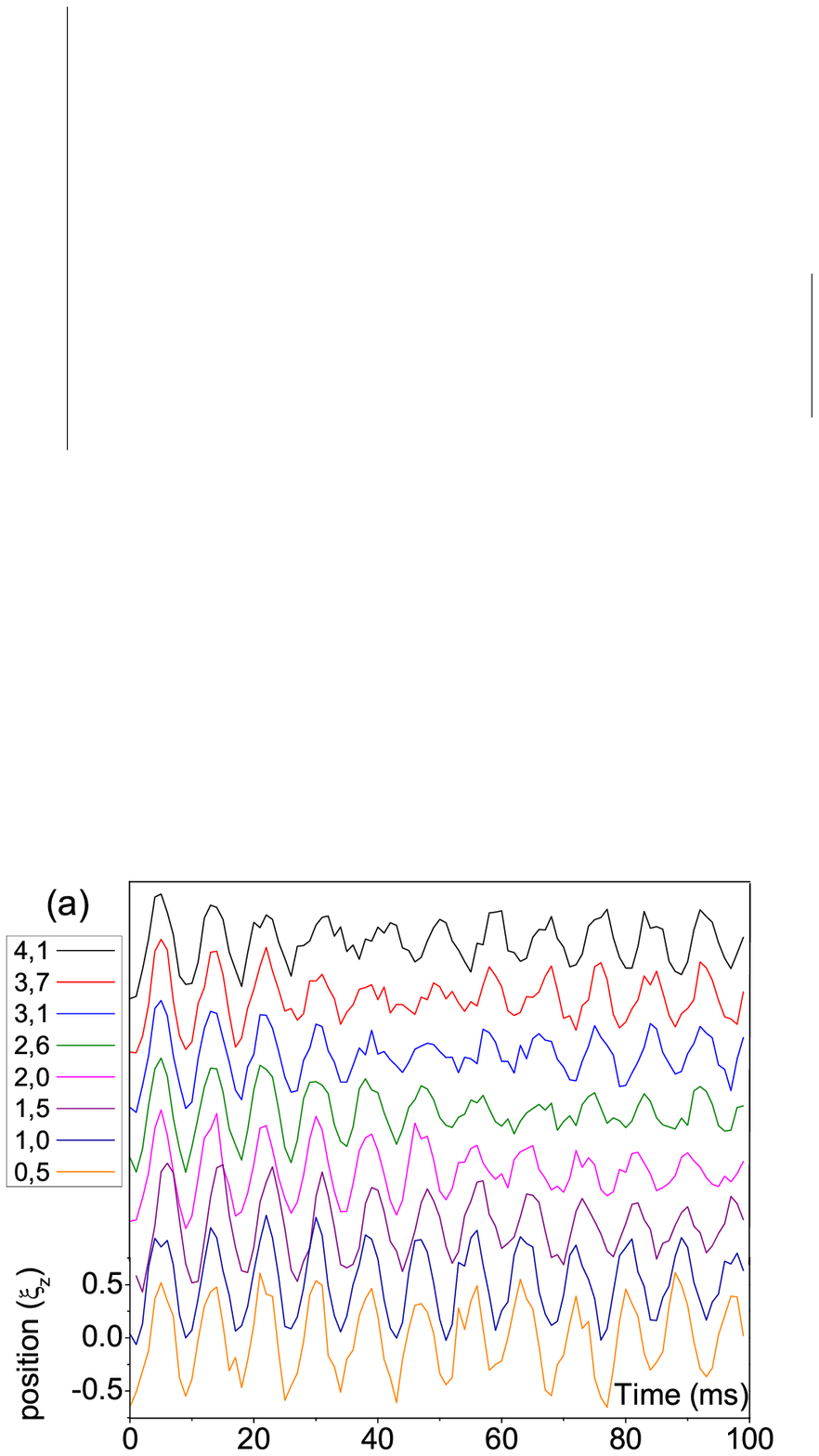}\\
\includegraphics[width=.97\linewidth,trim=-.17cm 0.8cm 0.6cm 1cm,clip]{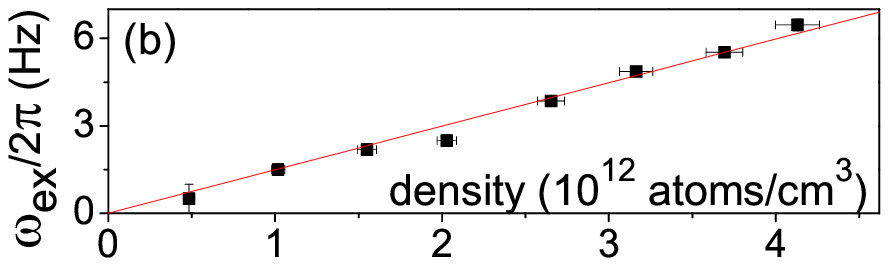}
\caption{(a) Center of mass position of the $\ket{\uparrow}$ component of the cold atom cloud versus time after a single $5\pi/2$ pulse for various atomic densities (in $10^{12}$ atoms/cm$^{3}$). Successive curves are offset vertically by $0.5\xi_z$. As the density increases, a beat appears between the trap frequency and the   increasingly faster identical spin rotation rate $\exRate$, characteristic for spin waves. (b) Fitting $\exRate$ versus density gives $2\pi\,1.4\,\text{Hz}/(10^{12} \text{atoms}/\text{cm}^3)\times \bar{n}.$} \label{fig:SpinWave}
\end{figure}
\end{center}

We initiate a spin-wave with a single $\tau=1.05$~ms excitation pulse of area $\Omega_0\tau=2.5\pi$. We use a multiple of $\pi/2$ to produce a larger spin inhomogeneity. This inhomogeneous spin population then oscillates in the trap and we observe the oscillation by holding the atoms in the trap for various times $t_h$ after the Rabi pulse, followed by 7\,ms of time-of-flight and state-selective absorption imaging.  Fig~\ref{fig:SpinWave}(a) shows the center of mass of the $\ket{\upp}$ component. The data for our lowest atom density exhibits a simple oscillation at $\omega_z$, and the $\ket{\dnn}$ cloud (not shown) oscillates out of phase. The center of mass of the total population shows no measurable oscillation. Increasing the density, we observe a collapse and revival of the oscillation at shorter and shorter times. At $t_h=80$\,ms, the oscillation for the highest density is out of phase with the lowest.  As we show below, this is a signature of a spin wave driven by ISR.

We can intuitively illustrate the spin dynamics by considering two localized atoms oscillating in a one-dimensional trap with frequency $\omega_z$ (Fig. \ref{Fig:BlochVectors}(b)). With the atoms initially on opposite sides of the trap center in state $\ket{\dnn}$, they are excited with a short Rabi pulse of mean area, typically $\theta_1=\Omega_0\tau=\pi/2$. Since $\Omega=\Omega(z)$, the atoms experience different Bloch vector rotations $\theta_1\pm\Delta\theta_1$. After the pulse, the atoms oscillate in the trap during the Ramsey interrogation time $T_R$ as in Fig. \ref{fig:SpinWave}(a). In the absence of interactions, each atom maintains its spin orientation, in a frame that rotates at the atomic transition frequency, and thus the spatial spin populations simply oscillate at $\omega_z$. Note that the phase of each atom's coherence is constant. However, if there are exchange interactions, the two spins will rotate with an ISR rate $\exRate$
 around their total spin as they repeatedly collide (Fig. \ref{Fig:BlochVectors}(b)) \cite{FuchsSpinWaves2002,TACCISRE2010}.
 In a time $\pi/\exRate$, the two atoms exchange their spin polarizations, producing a beat between $\exRate$ and $\omega_z$. This introduces the frequency $\exRate$ into the spatial oscillation of the spins, producing a beat between
$\exRate$ and $\omega_z$. For each curve, we extract $\exRate$, which varies linearly with density (Fig.~\ref{fig:SpinWave}(b)), as $2\pi\,1.4\,\text{Hz}/(10^{12} \text{at}/\text{cm}^3)\times \bar{n}$,  within our uncertainty. \footnote{This coefficient is five times smaller than $\exRateTh$ \cite{LaloeISRE1982}. Small amplitude dipolar spin waves oscillate at $\exRateTh/2$ \cite{Nikuni2002} and models show that this frequency can decrease by a factor 2 for large amplitude spin waves. Additionally, we experimentally observe $\exRate$ decreases by a factor of two as $\theta_1=2.2 \pi$ increases to $\theta_1=2.8 \pi$}


The spatiotemporal spin oscillation has important consequences for Ramsey spectroscopy. The second Ramsey pulse reads-out the phase of the atomic coherences. Between the pulses, the exchange interaction modulates the phase of each atom's coherence as the spins rotate about one another (Fig. \ref{Fig:BlochVectors}(b)). If we were to measure the transition probability of one of the atoms above \cite{GibbleShift2009}, the apparent resonance frequency would depend on $T_R$ - it would be modulated at $\exRate$. In the usual case when both atoms are detected, the frequency excursion of this modulated collision shift is reduced (Fig. \ref{Fig:BlochVectors}(c)). The shift averages to zero if the second Ramsey pulse is homogeneous. For an inhomogeneous second pulse, the two Bloch vectors experience different rotations $\theta_2\pm\Delta\theta_2=\Omega(z_i(T_R))\tau$ depending on their positions at the time of the pulse.
This gives them different weights in the Ramsey measurement, making the
cancellation incomplete, unless the second pulse is an odd multiple of
$\pi/2$, which reads out the phases of both atoms with the same
sensitivity \cite{GibbleShift2009}. This simple model predicts a clock shift
\begin{eqnarray}
\delta \nu = \frac{
\Delta\theta_1 \Delta\theta_2  \sin\left(\exRate T_R\right)
\cos\left(\omega_z T_R\right) \cos\theta_2}{4\pi T_R
\sin\theta_1\sin\theta_2}. \label{Eqn:deltanuRabiFort}
\end{eqnarray}
It extends the results in \cite{GibbleShift2009} to $\exRate T_R \ge1$ and unresolved sidebands. Here we linearize the dependence on $\Delta\theta_i$. A singlet-triplet basis provides helpful insight and also leads to Eq. (\ref{Eqn:deltanuRabiFort}). Before the first pulse the two atoms are in the triplet state $\ket{S,m_s}=\ket{1,-1}$. The inhomogeneous excitation pulse makes them partially distinguishable and populates the singlet state $\ket{0,0}$ \footnote{Only the singlet state for fermions can have s-wave collisions. For bosons with equal inter and intra-state scattering lengths, the triplet states all have the same energy shift and the singlet state has none. Therefore, taking out the common shift of the triplet states, such bosons have the same interactions as fermions, albeit with the opposite $\exRate$ \cite{FuchsSpinWaves2002,GibbleShift2009,GibbleViewpoint2010}.}, which then accrues a collisional phase shift during the Ramsey interrogation time \footnote{To derive Eq. \ref{Eqn:deltanuRabiFort} in the singlet-triplet basis, we consider short pulses, $\Omega_0\gg\omega_z,\exRate$, neglecting interactions during the pulses, and small inhomogeneities, $\delta\theta\ll 1$,
 exciting only a single trap sideband. We add the contributions for the upper and lower sidebands.}. 
When $\Omega_0\approx\omega_z,\exRate$, we numerically calculate the evolution of the $S=1$ pseudo-spin system, coherently including all transitions up to the 5th sideband. We also treat the 5\% scattering length difference $a_{\upp\upp}\lesssim a_{\upp\dnn}\lesssim a_{\dnn\dnn}$.

\begin{figure}
  \includegraphics[width=\linewidth,trim=0.2cm 0.8cm 0  0.5cm]{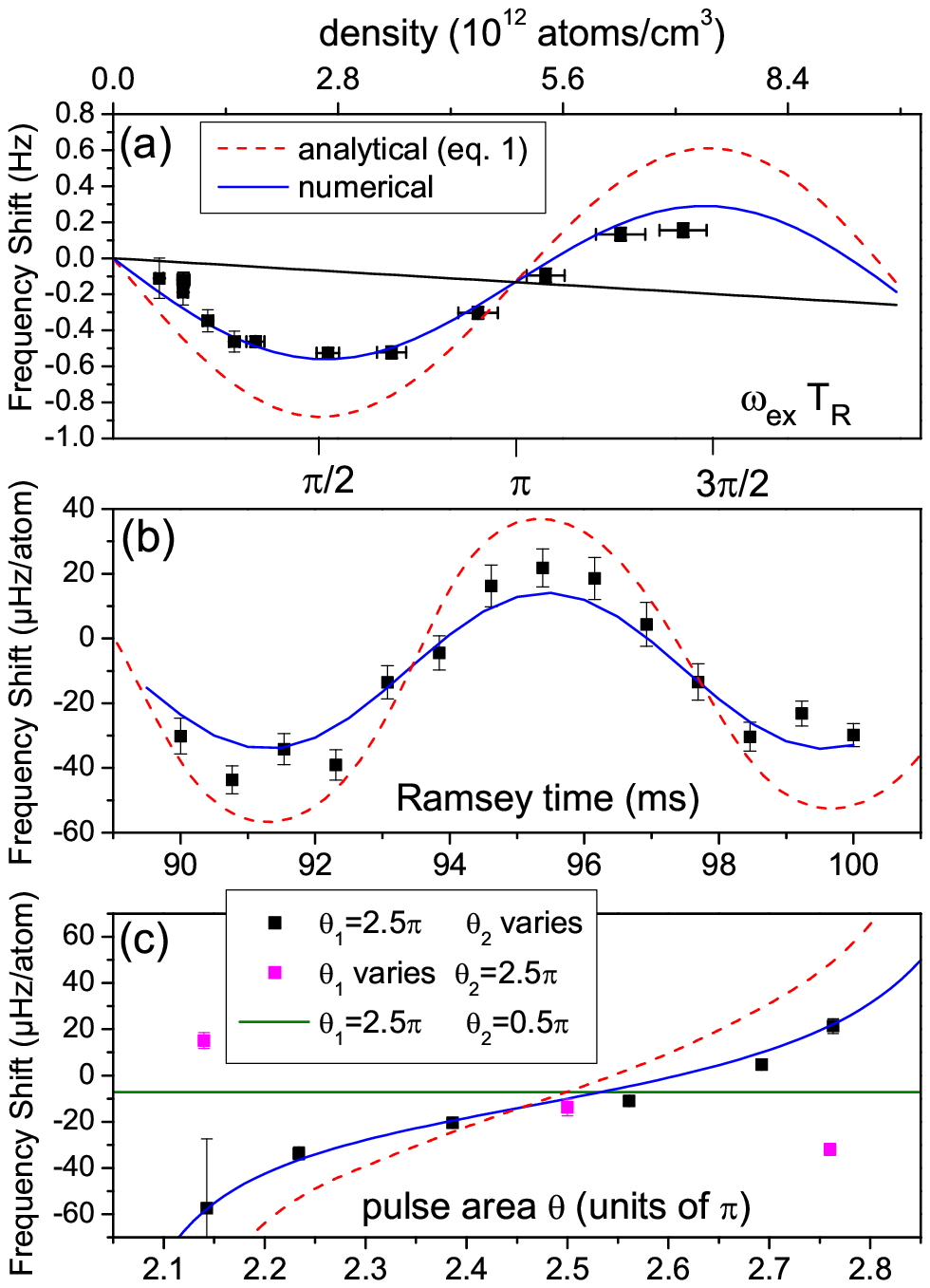}
  \caption{(a) Measured frequency shifts for a Ramsey sequence $T_R=100$~ms, $\theta_1=2.5\pi$ and $\theta_2=2.2\pi$ versus interaction strength, which is proportional to density. The shift is non-linear; in fact it oscillates as the density increases. For reference we plot the known shift for homogeneous excitations, $\delta\nu=(a_{\upp\upp}-a_{\dnn\dnn})/(2\pi a_{\upp\dnn})\exRate$ (black line).  (b) Density shift per atom $d\delta \nu/dN_{at}$ in the low interaction regime versus $T_R$ with $\theta_1=2.5\pi$ and $\theta_2= 2.2\pi$. The shift oscillates as the spin polarizations oscillate in the trap. (c)Dependence on the first and second pulse areas. \emph{Black squares}: $d\delta \nu/ dN_{at}$ for $T_R=92$~ms with $\theta_1=2.5\pi$ fixed and $\theta_2$ variable. The green horizontal line is a reference measurement with $\theta_2=\pi/2$ giving $-6\,\mu\mbox{Hz}/\mbox{atom}$, the expected frequency shift in the absence of ISR and predicted from the difference of scattering lengths. A numerical calculation based on the singlet-triplet model for two atoms (blue line) reproduces the data with no free parameters. Eq.~\ref{Eqn:deltanuRabiFort} qualitatively reproduces the observed behaviours, but overestimates the shift by 60\% (red line). \emph{Magenta dots}: $\theta_1$ variable and $\theta_2=2.5\pi$ fixed. This dependence
is not reproduced by either of the models.  \label{fig:shifts}}
\end{figure}

To experimentally test eq. (\ref{Eqn:deltanuRabiFort}), we measure the shift of the clock's frequency with a Ramsey sequence for the same range of densities as in Fig.~\ref{fig:SpinWave}. Fig.~\ref{fig:shifts}(a) shows the measured shift as a function of density for two $\tau_{1,2}=1.05$\,ms pulses separated by a $T_R=100$\,ms interrogation time, which is close to a multiple of the trap period. The first pulse area is  $\theta_1=5\pi/2$ and the second is $\theta_2=2.2 \pi$. Like above, a large pulse area is used to increase
the inhomogeneity. The observed frequency shift indeed oscillates as a function of density, giving a frequency shift that is inconsistent with the often-used mean-field shift \cite{Harber2002,Zwierlein2003,Campbell2009,ReyFermionShift2009}. Moreover, the first zero of $\delta\nu$ indeed occurs for $\exRate T_R=\pi$, with the value of $\exRate$ being determined from the data in Fig.~\ref{fig:SpinWave} for this density. This confirms a distinguishing prediction of eq. (\ref{Eqn:deltanuRabiFort}).


We also measure the shift as a function of $T_R$ (Fig.~\ref{fig:shifts}(b)). Again we use $\theta_1= 2.5 \pi$ and $\theta_2=2.2 \pi$ and determine the frequency shift $\delta\nu$ for each of the atomic densities, $\overline{n}$ $\approx$ $\{0.4, 0.8, 1.3, 1.7 \} \,10^{12}$ at/cm$^{3}$. From a linear fit we extract the slope $\alpha=d\delta \nu/dN_{at}$. This suppresses potential density-independent frequency shifts that vary with $T_R$.  Whenever the spatial spin distribution is the same for the first and second Ramsey pulses, the shift has the same sign and has the opposite sign when the spatial spin distribution reverses.
This emphasizes the importance of the correlation between the inhomogeneities of the first and second Ramsey pulses. In the low-density regime, ($\overline{n}\lesssim 10^{12}\mbox{at}/\mbox{cm}^{3}$), Eq. \ref{Eqn:deltanuRabiFort} predicts that $\alpha$ in Fig. \ref{fig:shifts}(b) should oscillate at $\omega_z$. Here, the mean frequency shift is offset, as expected from the small scattering length differences.

A distinguishing feature of the s-wave fermion collisions (eq.~\ref{Eqn:deltanuRabiFort}) is that the frequency shift depends on the \emph{area} of the second Ramsey pulse \cite{GibbleShift2009}. For homogeneous excitations of ultracold bosons, the frequency shift depends on the area of the first Ramsey pulse, which determines the population difference of the clock states for the collisions during the Ramsey interrogation time. This unique feature was not demonstrated in the observations of collisional shifts of lattice clocks \cite {Campbell2009,ReyFermionShift2009,Yu2010,SwallowsSupprCollShift2011, Bishof2011,Ludlow2011}. We apply $\theta_1=2.5\pi$, $T_R=92$~ ms and vary the second pulse area $\theta_2$ via the microwave power, keeping the duration fixed. The resulting $\alpha$ is shown in Fig.~\ref{fig:shifts}(c) (black squares). The $1/\tan\theta_2$ dependence predicted in \cite{GibbleShift2009} and (\ref{Eqn:deltanuRabiFort}) is clearly visible.

To show the quantitative agreement between the data of Fig. \ref{fig:shifts} and our models we use the experimental parameters, including $\delta_1$ which is independently determined from the resonant Rabi flopping with the respective model (Fig. \ref{Fig:BlochVectors}(a)).The numerical model reproduces the data \footnote{To extract $\alpha$ from the numerical model, we calculate the shift for the 4 densities and extract $\alpha$ from a linear fit, in the same way that we analyze the data.}. The analytical model eq. \ref{Eqn:deltanuRabiFort} reproduces the oscillations but overestimates the amplitude of the shift by 60\%. Including many sidebands, the numerical model gives better agreement as expected.

We also vary the area of the first Ramsey pulse $\theta_1$, keeping $\theta_2=5\pi/2$ fixed. Surprisingly, the shift is comparable to when $\theta_2$ is varied. Eq. (\ref{Eqn:deltanuRabiFort}) has a small dependence on $\theta_1$ when $\theta_2$ is not exactly $5\pi/2$ but the predicted shift is much smaller than observed and would not change sign near $\theta_1=2.5\pi$. Similarly, shifts due to the small difference of scattering lengths $a_{\upp\upp}+a_{\dnn\dnn}-2a_{\upp   \dnn}$ are too small. However, we note that our first pulse motionally excites the spin components beyond a simple dipolar excitation. This leads to an oscillation of the cloud size, which unexpectedly varies with $\theta_1$. We speculate that this may produce some dephasing \footnote{Such oscillations could create additional spatial inhomogeneities of the transition frequency through the trap, and add a new source of dephasing. Any dephasing directly couples  pair-wise singlet and triplet states during the Ramsey time and generally leads to a dependence of the measured frequency on $\theta_1$. Since the contrast does not decrease significantly, the dephasing rate would have to be less than $\exRate$}, and thereby leads to a dependence on $\theta_1$.


The striking connection between spin-waves and s-wave fermion collision shifts demonstrated here is very general. We can elucidate this connection by considering the resolved sideband regime, used in many ultraprecise
atomic clocks, including optical lattice clocks. With resolved sidebands and weak interactions, even though the clock field cannot change the
motional state of the atoms, we show that spin waves are excited. Here, the spatial inhomogeneity may
give a low energy atom in vibrational state $\ket{\alpha}$ a large
pulse area and a high energy atom in $\ket{\beta}$ a small pulse area,
directly populating pair-wise singlet states $\ket{0,0}$
\cite{GibbleShift2009}. After the pulse, the two fermionic atoms evolve as
$\ket{\Psi(T)}= s e^{i \omega_{ex} T} \ket{0,0} \{\ket{\alpha\beta}\}^+
+(t\ket{1,0}+u\ket{1,1}+d\ket{1,-1}) \{\ket{\alpha\beta}\}^-$, where
$\ket{1,m_S}$ are triplet states, $\{\}^{(-)+}$ denotes
(anti)symmetrization, and $u, d, t$, and $s$ are the state
amplitudes. Rewriting this two particle wavefunction as
$\ket{\Psi(T)} = 1/\sqrt{2}(t+s e^{i \omega_{ex} T})\{\ket{\upp\alpha}
\ket{\dnn\beta}\}^- +1/\sqrt{2}(t-s e^{i \omega_{ex} T})
\{\ket{\dnn\alpha} \ket{\upp\beta}\}^- +(u\ket{1,1}
+d\ket{1,-1})\{\ket{\alpha\beta}\}^-$, we see that the $\ket{\upp}$
populations in state $\ket{\alpha}$ and $\ket{\beta}$ have an
explicit oscillation at $\omega_{ex}$ -- at different times T, the
$\ket{\upp}$ population in the vibrational states are different. Thus, whenever there is a  fermion collision shift, spin waves must also exist, and the fermion collision shift will oscillate as the spin populations oscillate in the trap.

In summary, we observe characteristic behaviors of the collisional frequency shifts due to inhomogeneous excitations in an atomic clock. The inhomogeneous excitations create spin-waves, which we show are inextricably connected to the s-wave frequency shifts of fermion clocks, including optical-frequency lattice clocks. We directly excite dipolar spin waves via an amplitude gradient of the excitation field. The spin populations oscillate, exhibiting a beat between the trap frequency and the frequency of spin rotation due to particle interactions. This leads to a collisional frequency shift that oscillates as the spin populations oscillate in the trap. We observe that the clock collision shift does not vary linearly with the atomic density and, in the spin-wave regime, varying the Ramsey interrogation time $T_R$ (Fig.~\ref{fig:shifts}(b)) could help to evaluate the accuracy of atomic clocks. The frequency shift exhibits the novel dependence on the area of the second Ramsey pulse, in stark contrast to the mean field expressions for frequency shifts with homogeneous excitations \cite{GibbleShift2009}. While we intentionally exaggerate the spin wave excitations here, these frequency shifts can be minimized by using spatially homogenous fields, using sideband resolved pulses, and avoiding the Knudsen regime so that trap-state changing collisions further suppress the fermion shift.




We acknowledge contribution of F. Reinhard. This work was supported by the Institut Francilien pour la Recherche sur les Atomes Froids (IFRAF), by the ANR (grant ANR-09-NANO-039) and by EU through the AQUTE Integrated Project (grant agreement 247687) and the project EMRP IND14, the NSF, and Penn State.

%

\end{document}